\documentclass[aps,prb,twocolumn,preprintnumbers,amsmath,amssymb,superscriptaddress]{revtex4-2}%

\usepackage{amsmath}
\usepackage{graphicx}
\usepackage{bm}
\usepackage{multirow}

\begin{document}

\title{Multiple Magnetic Transitions in the Trilayer Nickelate Pr$_4$Ni$_3$O$_{10}$ Revealed by Muon-Spin Rotation}

\author{Rustem Khasanov}
 \email{rustem.khasanov@psi.ch}
 \affiliation{PSI Center for Neutron and Muon Sciences CNM, 5232 Villigen PSI, Switzerland}

\author{Thomas J. Hicken}
 \affiliation{PSI Center for Neutron and Muon Sciences CNM, 5232 Villigen PSI, Switzerland}

\author{Zurab Guguchia}
 \affiliation{PSI Center for Neutron and Muon Sciences CNM, 5232 Villigen PSI, Switzerland}

\author{Shangxiong Huangfu}
 \affiliation{Department of Physics, University of Zurich, Winterthurerstrasse 190, CH-8057 Zurich, Switzerland}
 \affiliation{Key Laboratory of Advanced Functional Materials, Ministry of Education, College of Materials Science and Engineering, Beijing University of Technology, Beijing 100124, China}

\author{Hubertus Luetkens}
 \affiliation{PSI Center for Neutron and Muon Sciences CNM, 5232 Villigen PSI, Switzerland}

\author{Ekaterina Pomjakushina}
 \affiliation{PSI Center for Neutron and Muon Sciences CNM, 5232 Villigen PSI, Switzerland}

\author{Vladimir Pomjakushin}
 \affiliation{PSI Center for Neutron and Muon Sciences CNM, 5232 Villigen PSI, Switzerland}

\author{Andreas Schilling}
 \affiliation{Department of Physics, University of Zurich, Winterthurerstrasse 190, CH-8057 Zurich, Switzerland}

\author{Igor Plokhikh}
 \affiliation{PSI Center for Neutron and Muon Sciences CNM, 5232 Villigen PSI, Switzerland}
 \affiliation{TU Dortmund University, Department of Physics, Dortmund, 44227, Germany}

\author{Dariusz J. Gawryluk}
 \affiliation{PSI Center for Neutron and Muon Sciences CNM, 5232 Villigen PSI, Switzerland}

\date{\today}

\begin{abstract}
A muon-spin rotation/relaxation ($\mu$SR) study of the trilayer Ruddlesden--Popper nickelate Pr$_4$Ni$_3$O$_{10}$ was performed at ambient pressure and under hydrostatic pressure up to 2.2~GPa.
Three magnetic transitions were identified at ambient pressure: the onset of spin-density-wave (SDW) order at $T_{\rm SDW} \simeq 158$~K, an intermediate-temperature transition at $T^{\ast} \simeq 90$--100~K, and a low-temperature transition at $T_{\rm SDW}^{\rm Pr} \simeq 25$--27~K. While the intermediate transition at $T^{\ast}$ induces only minor changes in the internal-field distribution, the transition at $T_{\rm SDW}^{\rm Pr}$ is accompanied by a pronounced reconstruction of the magnetic structure, consistent with previous reports attributing enhanced interlayer coherence to the ordering of the Pr sublattice.
The high-temperature transition at $T_{\rm SDW}$ is characterized by the sharp development of static internal magnetic fields with a narrow transition width of $0.65(4)$~K. Weak-transverse-field measurements reveal a finite thermal hysteresis of $0.27(6)$~K, with $T_{\rm SDW}^{\rm warming} > T_{\rm SDW}^{\rm cooling}$, indicating weakly first-order-like behavior. Hydrostatic pressure suppresses $T_{\rm SDW}$ linearly and reduces the ordered Ni magnetic moment $M$, with corresponding rates of ${\rm d}T_{\rm SDW}/{\rm d}p = -4.9(1)$~K/GPa and ${\rm d}\ln M/{\rm d}p = -2.0(5)\times10^{-2}$~GPa$^{-1}$, respectively, thereby demonstrating a gradual weakening of the spin-density-wave instability under compression.
\end{abstract}

\maketitle

\section{Introduction}

The discovery of superconductivity in layered nickelates has revived strong interest in the broader family of Ruddlesden--Popper (RP) nickelates as platforms for correlated-electron phenomena \cite{Li_Nature_2019, Zeng_PRL_2020, Lee_APLMater_2020, Osada_NanoLett_2020, Lee_Nature_2023, Puphal_NatRevPhys_2026, Sun_Nature_2023, Zhang_NatCom_2020, Zhu_Nature_2024, Sakakibara_PRB_2024, Wang_CPL_2024, Li_CPL_2024, Wang_PRX_2024, Zhang_arxiv_2023, Wu_PRB_2001, Xu_NatCom_2025, Huangfu_PRR_2020, Li_NatCom_2017, Zhang_PRM_2020, Khasanov_La327_NatPhys_2024, Wang_Nature_2024, Shi_arxiv_2025, Li_arxiv_2025, Pei_arxiv_2024, Zhang_arxiv_2025, Huangfu_PRB_2020, Huangfu_PRB_2020_2, Carvalho_JAP_2000, Seo_InorgChem_1996, Li_SciChina_2024, Huang_Arxiv_2025, Khasanov_LaPr327-pressure_PRR_2025}.
In particular, the recent observation of superconductivity in bilayer and trilayer RP nickelates under pressure has triggered intensive efforts to clarify the interplay between density-wave order, magnetism, and superconductivity in these systems \cite{Sun_Nature_2023, Wang_CPL_2024,  Li_CPL_2024, Wang_PRX_2024, Wang_Nature_2024, Zhu_Nature_2024, Sakakibara_PRB_2024, Zhang_arxiv_2023, Huang_Arxiv_2025}.

The RP series R$_{n+1}$Ni$_n$O$_{3n+1}$ provides a unique opportunity to systematically tune electronic dimensionality through the number of NiO$_2$ layers $n$. By increasing $n$, the strength of interlayer coupling, the electronic bandwidth, and the balance between competing ordered states can be continuously modified \cite{Zhang_NatCom_2020, Wu_PRB_2001, Wang_Nature_2024, Pei_arxiv_2024, Zhang_arxiv_2025, Huangfu_PRB_2020, Huangfu_PRB_2020_2, Huangfu_PRR_2020, Li_NatCom_2017, Zhang_PRM_2020, Carvalho_JAP_2000, Seo_InorgChem_1996, Huang_Arxiv_2025, Khasanov_La327-1313_Arxiv_2025, Khasanov_La4310_PRR_2026}.
This structural tunability makes RP nickelates a natural laboratory for investigating how magnetic and density-wave instabilities evolve with dimensionality and how their suppression relates to the emergence of superconductivity.

Among these materials, trilayer RP nickelates ($n=3$) have attracted particular attention. Previous studies have established that compounds such as La$_4$Ni$_3$O$_{10}$ and Pr$_4$Ni$_3$O$_{10}$ host coexisting charge-density-wave (CDW) and spin-density-wave (SDW) orders \cite{Zhang_NatCom_2020, Samarakoon_PRX_2023, Jia_PRX_2026}.
In these systems, the magnetically ordered state develops within a charge-ordered background, reflecting strong coupling between spin, charge, and lattice degrees of freedom. Such intertwined density-wave behavior is reminiscent of closely related correlated oxides, including cuprates \cite{Tranquada_Nature_1995, Ghiringhelli_Sciense_2012, Chang_NatPhys_2012, Fradkin_RMP_2015, Keimer_Nature_2015, Kivelson_RMP_2003, Guguchia_PRL_2020},
and hole-doped nickelates \cite{Ricci_PRL_2021, Zhang_PNAS_2016, Zhang_PRL_2019},
where competing or intertwined SDW and CDW orders are known to play a central role in shaping the phase diagram.
Importantly, superconductivity emerges only after substantial (or even complete) suppression of the intertwined charge- and spin-density-wave state under applied pressure \cite{Zhu_Nature_2024, Zhang_arxiv_2023, Xu_NatCom_2025, Pei_arxiv_2024, Huang_CPL_2024, Pei_JACS_2026}.
This sequence indicates that the coupled CDW/SDW phase forms the normal-state background from which superconductivity develops, underscoring the importance of establishing a detailed microscopic understanding of the intertwined order and its evolution under compression.

Pr$_4$Ni$_3$O$_{10}$ occupies a particularly interesting position within the trilayer family. Its crystal structure consists of three NiO$_2$ planes forming a trilayer block separated by rock-salt-type layers \cite{Pei_arxiv_2024, Huangfu_PRB_2020, Zhang_PRM_2020, Zhang_arxiv_2025, Samarakoon_PRX_2023, Tsai_JSSC_2020, Pei_JACS_2026}.
This geometry implies strong magnetic and electronic coupling within each trilayer unit, while the coupling between adjacent blocks is substantially weaker \cite{Zhang_NatCom_2020, Samarakoon_PRX_2023, Khasanov_La327-1313_Arxiv_2025, Khasanov_La4310_PRR_2026}.
Such a hierarchy of interactions promotes anisotropic, quasi-two-dimensional behavior, which can enhance fluctuation effects and favor intertwined spin- and charge-ordered states \cite{Zhang_NatCom_2020, Samarakoon_PRX_2023, Jia_PRX_2026}.
In addition, the presence of the Pr sublattice, in contrast to the nonmagnetic La ions in La$_4$Ni$_3$O$_{10}$, introduces additional magnetic degrees of freedom that may enhance three-dimensional magnetic coupling and further enrich the low-temperature magnetic phase diagram \cite{Samarakoon_PRX_2023}.

Despite growing experimental activity, the microscopic nature of magnetism in Pr$_4$Ni$_3$O$_{10}$ remains unresolved. In particular, several key questions remain open: (i) what is the detailed sequence of magnetic transitions upon cooling, (ii) what is the character of the primary high-temperature SDW transition and does it exhibit signatures of first-order behavior, and (iii) how robust is the static Ni magnetic order under hydrostatic pressure in the regime preceding superconductivity? Addressing these issues is essential for establishing a quantitative magnetic phase diagram and for clarifying how the suppression of magnetism under pressure connects to the eventual emergence of superconductivity in trilayer RP nickelates.

Muon-spin rotation/relaxation ($\mu$SR) is a powerful local probe ideally suited to address these questions. As a microscopic technique sensitive to internal magnetic fields at the muon stopping site, $\mu$SR allows direct detection of static magnetism, quantitative determination of magnetic volume fractions, and precise tracking of ordered moments across phase transitions. Importantly, it provides access to magnetic order parameters and phase coexistence without relying on macroscopic transport or bulk thermodynamic signatures, which can be influenced by sample inhomogeneity or percolative effects.

In this work, we present a comprehensive $\mu$SR investigation of Pr$_4$Ni$_3$O$_{10}$ at ambient pressure and under hydrostatic pressure up to 2.2~GPa, i.e., in the regime preceding the onset of superconductivity reported at substantially higher pressures. We establish the sequence of magnetic transitions, determine the nature of the primary high-temperature SDW transition, and quantify the pressure evolution of both the ordering temperature and the ordered Ni magnetic moment.

Our results reveal that the high-temperature SDW transition is sharp and exhibits a finite thermal hysteresis, indicating weakly first-order--like behavior. Two additional magnetic transitions occur at lower temperatures: one associated with a subtle modification of the SDW state and another accompanied by a pronounced reconstruction of the magnetic field distribution consistent with the involvement of the Pr sublattice. Under applied pressure, the SDW transition temperature decreases linearly, while the ordered Ni moment is gradually reduced. These findings provide a quantitative magnetic phase diagram of Pr$_4$Ni$_3$O$_{10}$ in the normal state and establish the magnetic baseline from which superconductivity eventually emerges in trilayer RP nickelates.

The paper is organized as follows. Section~\ref{sec:Experimental_details} presents the experimental details and describes the sample preparation procedure, as well as its characterization by means of neutron powder diffraction and thermogravimetric analysis (Sec.~\ref{sec:Sample_preparation}). It also provides a brief description of the muon-spin rotation/relaxation experiments (Sec.~\ref{sec:MuSR-experiments}) and the $\mu$SR data analysis procedure (Sec.~\ref{sec:MuSR-data_analysis}). Section~\ref{sec:Results} presents the results obtained from ambient-pressure (Sec.~\ref{sec:muSR_ambient-pressure}) and high-pressure (Sec.~\ref{sec:WTF-ZF-pressure}) $\mu$SR studies performed in zero- and weak transverse-field. The experimental data are discussed in Sec.~\ref{sec:Discussions}, and the conclusions are presented in Sec.~\ref{sec:Conclusions}.

\section{Experimental details}\label{sec:Experimental_details}

\subsection{Sample preparation and characterisation}\label{sec:Sample_preparation}

The polycrystalline Pr$_4$Ni$_3$O$_{10}$ compound was synthesized using a modified sol--gel Pechini method with 2-hydroxypropane-1,2,3-tricarboxylic acid as the chelating agent, followed by a solid-state reaction, as described in Ref.~\cite{Huangfu_PRB_2020}. A portion of this sample, used for hysteresis studies in weak-transverse-field $\mu$SR experiments, underwent an additional annealing treatment in oxygen at a pressure of $\simeq 1.6$~bar at room temperature, which increased to $\simeq 2.5$~bar at 1000$^\circ$C. The sample was kept at 1000$^\circ$C for 6~h, followed by a post-annealing step at 500$^\circ$C for 24~h. This entire annealing procedure was repeated four times.

\begin{figure*}[htb]
\centering
\includegraphics[width=1.0\linewidth]{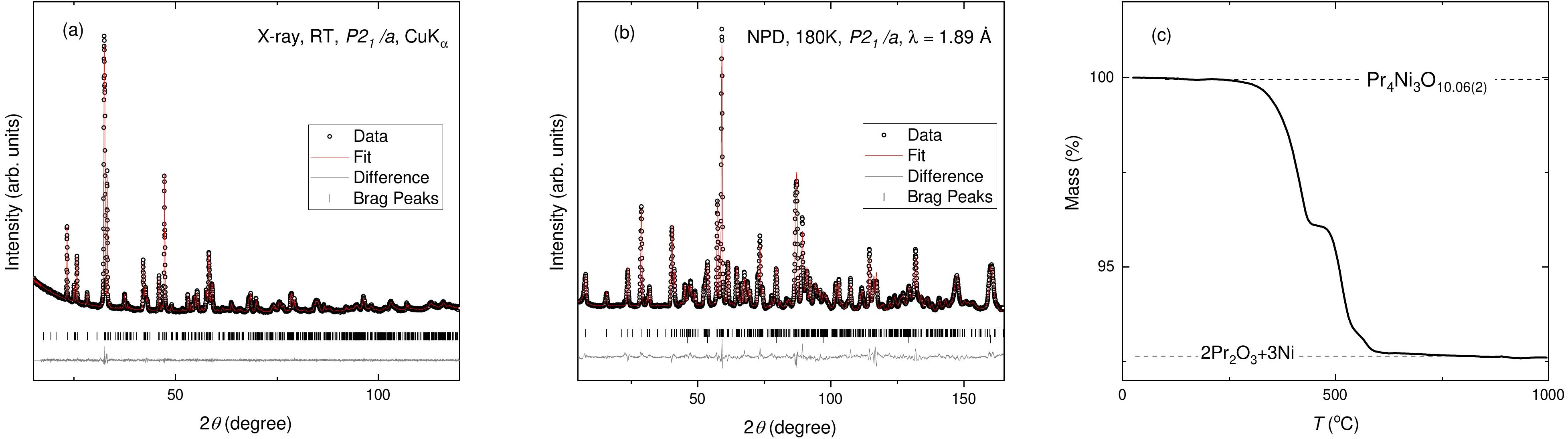}
\caption{
(a) Room-temperature X-ray powder diffraction pattern of Pr$_4$Ni$_3$O$_{10}$ measured in Bragg--Brentano geometry using Cu K$\alpha$ radiation. Black circles represent the experimental data, and the red solid line shows the Le Bail refinement using the monoclinic $P2_1/a$ space group (No.~14).
(b) Neutron powder diffraction pattern collected at $T = 180$~K using the high-resolution thermal neutron diffractometer HRPT \cite{HRPT} with a wavelength of $\lambda = 1.89$~\AA\ in the $2\theta$ range $3.55^\circ$--$164.50^\circ$ with a step size of $0.05^\circ$. Black circles represent the experimental data, the red solid line shows the Rietveld refinement using the monoclinic $P2_1/a$ space group (No.~14).
The ticks and the grey curves in (a) and (b) indicate Bragg reflection positions, and the difference between experimental and calculated intensities, respectively.
(c) Thermogravimetric analysis  of Pr$_4$Ni$_3$O$_{10}$ measured during hydrogen reduction (7~vol.\% H$_2$ in He) with a heating rate of 1~$^\circ$C/min. The total mass loss corresponds to the reduction of fully oxidized Pr$_4$Ni$_3$O$_{10.08(2)}$ to metallic Ni and Pr$_2$O$_3$.
}
\label{fig:Neutron-TG}
\end{figure*}

Phase purity was confirmed by X-ray and neutron powder diffraction (NPD) experiments. Laboratory X-ray diffraction was performed at room temperature in Bragg--Brentano geometry using a Bruker AXS D8 Advance diffractometer (Bruker AXS GmbH, Karlsruhe, Germany) equipped with Ni-filtered Cu K$\alpha$ radiation and a 1D LynxEye position-sensitive detector, see Fig.~\ref{fig:Neutron-TG}~(a). Le Bail refinement of the diffraction pattern was carried out using the FullProf Suite \cite{RodriguezCarvajal_PhysB_1993, RodriguezCarvajal_ActaCrystB_2025, Roisnel_MSF_2001, Arcelus_JApplCryst_2024}. The refinement yielded the following unit-cell parameters for the monoclinic structure (space group $P2_1/a$, No.~14): $a = 5.3767(1)~\text{\AA}$, $b = 5.4645(1)~\text{\AA}$, $c = 14.0247(1)~\text{\AA}$, and $\beta = 100.78(1)^\circ$, consistent with literature data \cite{Zhang_PRM_2020}.

The NPD data were collected at $T = 180$~K using the high-resolution thermal neutron powder diffractometer HRPT \cite{HRPT} with a neutron wavelength of $\lambda = 1.89$~\AA, see Fig.~\ref{fig:Neutron-TG}(b). Diffraction patterns were recorded over a $2\theta$ range of $3.55^\circ$--$164.50^\circ$ with a step size of $0.05^\circ$. The monoclinic $P2_1/a$ (No.~14) structure reported in Ref.~\cite{Zhang_PRM_2020} was used as the initial model and refined via the Rietveld method using the \textsc{Jana2020} software package \cite{Jana1, Jana2}. A minor NiO impurity ($\sim 1.8\%$) was included in the refinement. The final refinement yielded unit-cell parameters of $a = 5.3703(2)$~\AA, $b = 5.4634(2)$~\AA, $c = 14.0004(6)$~\AA, and $\beta = 107.74(2)^\circ$, with oxygen occupancies refined and confirmed to be unity. The refined neutron structural parameters, including atomic positions and isotropic displacement parameters, are summarized in Table~\ref{tab:NPD}.

\begin{table}[htbp]
\centering
\caption{Atomic positions and isotropic displacement parameters.}
\label{tab:NPD}
\begin{tabular}{| l | c | c | c | c | c | c |}
\hline
Atom & Position & Occ. & $x/a$ & $y/b$ & $z/c$ & $U_{\mathrm{iso}}$ \\
\hline
Pr1 & 4e & 1 & 0.694(1) & 0.501(1) & 0.9020(3) & \multirow{2}{*}{0.0058(6)} \\
Pr2 & 4e & 1 & 0.589(1) & 0.494(1) & 0.6333(4) &  \\ \hline

Ni1 & 2b & 1 & 0.5 & 0 & 0.5 & \multirow{2}{*}{0.00535(4)} \\
Ni2 & 4e & 1 & 0.6411(6) & 0.0010(8) & 0.7793(1) &  \\ \hline

O1 & 4e & 1 & 0.719(1) & 0.9477(8) & 0.9310(3) & \multirow{5}{*}{0.0118(1)} \\
O2 & 4e & 1 & 0.432(1) & 0.9374(7) & 0.3613(3) &  \\
O3 & 4e & 1 & 0.895(1) & 0.757(2) & 0.7661(3) &  \\
O4 & 4e & 1 & 0.899(1) & 0.2482(2) & 0.7962(3) &  \\
O5 & 4e & 1 & 0.281(1) & 0.726(1) & 0.5211(4) &  \\ \hline
\end{tabular}
\end{table}

Thermogravimetric analysis  of hydrogen reduction was performed to determine the oxygen content of the studied sample, see Fig.~\ref{fig:Neutron-TG}(c). Measurements were carried out using a NETZSCH 449F1 simultaneous thermal analyzer. The initial sample mass was 88.20~mg. The reducing gas consisted of 7~vol.\% hydrogen (Messer Schweiz AG, 5N) in helium (PanGas, 6N). The sample was heated from room temperature to 1000~$^\circ$C at a rate of 1~$^\circ$C/min and subsequently cooled to room temperature. The methodology used to determine the oxygen content and estimate associated uncertainties is described elsewhere \cite{Plokhikh_arXiv_2025}. The total observed weight loss corresponds to the reduction of fully oxidized Pr$_4$Ni$_3$O$_{10.08(2)}$ to metallic nickel and Pr$_2$O$_3$.

\subsection{$\mu$SR experiments}\label{sec:MuSR-experiments}

Ambient-pressure and high-pressure muon-spin rotation/relaxation ($\mu$SR) experiments were performed at the $\pi$M3 and $\mu$E1 beamlines of the Paul Scherrer Institute (PSI, Villigen, Switzerland) using the GPS (General Purpose Surface, Ref.~\cite{Amato_RSI_2017}) and GPD (General Purpose Decay, Refs.~\cite{Khasanov_HPR_2016, Khasanov_JAP_2022}) spectrometers.
Quasi-hydrostatic pressures up to $\simeq 2.2$~GPa were generated using double-wall piston-cylinder clamp cells made of nonmagnetic MP35N alloy \cite{Khasanov_HPR_2016}. The $\mu$SR data were analyzed using the MUSRFIT software package \cite{MUSRFIT}.

The measurements were carried out in two complementary experimental configurations: zero-field (ZF) and weak transverse-field (WTF):
\paragraph*{Zero-field  $\mu$SR.}
In the ZF configuration, no external magnetic field was applied, so that the implanted muon spins probed exclusively the internal magnetic fields generated by static or fluctuating electronic moments in the sample. ZF-$\mu$SR measurements therefore provide direct access to the magnitude and distribution of internal magnetic fields at the muon stopping sites. Since these internal fields are proportional to the ordered magnetic moments, ZF-$\mu$SR enables quantitative characterization of the magnetic order parameter and its temperature dependence \cite{Amato-Morenzoni_book_2024, Yaouanc_book_2011, Blundell_book_2022}.
\paragraph*{Weak transverse-field  $\mu$SR.}
In the WTF configuration, a small magnetic field was applied perpendicular to the initial muon-spin polarization. WTF-$\mu$SR measurements were primarily used to determine the temperature evolution of the magnetic volume fraction. Muons stopping in paramagnetic (nonmagnetic) regions precess coherently in the applied field, producing a well-defined oscillatory asymmetry signal. In contrast, muons implanted in magnetically ordered regions experience large and broadly distributed internal magnetic fields, leading to rapid depolarization and suppression of the oscillating WTF signal. The relative amplitude of the oscillatory component therefore provides a direct measure of the nonmagnetic fraction of the sample \cite{Amato-Morenzoni_book_2024, Yaouanc_book_2011, Blundell_book_2022}.

\begin{figure*}[htb]
\centering
\includegraphics[width=1.0\linewidth]{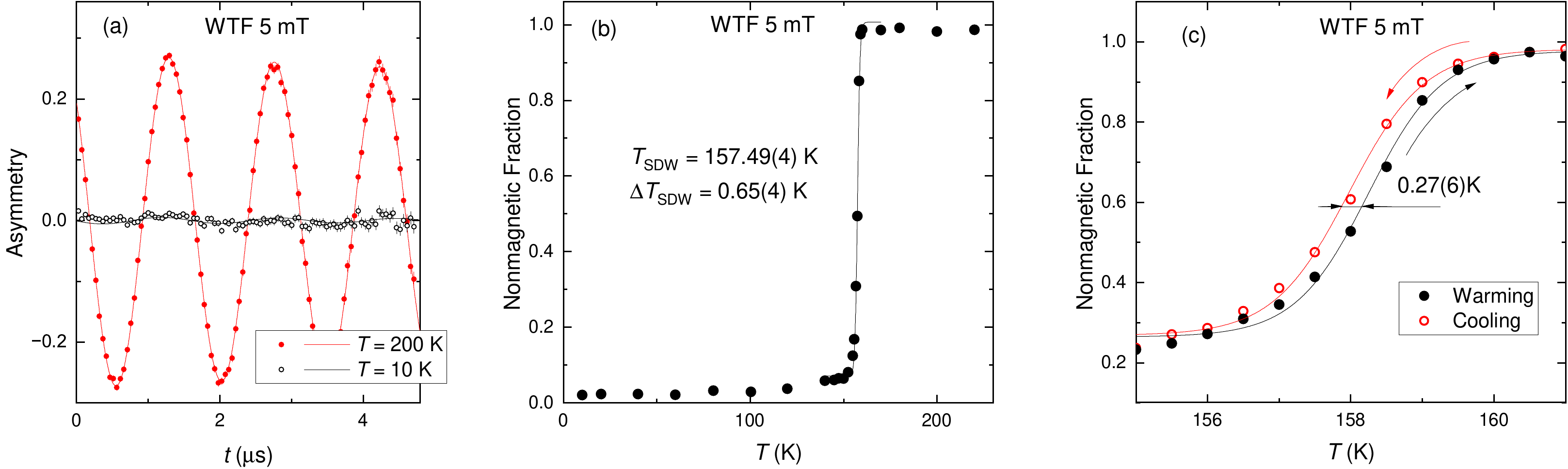}
\caption{
(a) WTF-$\mu$SR time spectra measured at $B_{\rm WTF}=5$~mT at $T=10$~K and 200~K. The solid lines represent fits using Eq.~\ref{eq:WTF}.
(b) Temperature dependence of the nonmagnetic volume fraction $1-f_{\rm m}$. The solid line represents a fit of $(1-f_{\rm m})$ vs.\ $T$ in the vicinity of the SDW transition.
(c) Temperature dependence of $1-f_{\rm m}$ measured upon warming (black symbols) and cooling (red symbols). The solid lines represent fits using Eq.~\ref{eq:WTF}. The sample used in hysteresis experiments underwent additional annealing (see Sec.~\ref{sec:Sample_preparation}), which may have resulted in a slightly different oxygen content ($\delta$) in Pr$_4$Ni$_3$O$_{10+\delta}$. Given the sensitivity of density-wave transitions to oxygen stoichiometry, such small variations in $\delta$ can account for the minor shift in $T_{\rm SDW}$. The larger offset likely reflects a small residual nonmagnetic fraction and/or a slightly different background contribution.
}
\label{fig:WTF_ZP}
\end{figure*}

\subsection{ZF- and WTF-$\mu$SR data analysis procedure}\label{sec:MuSR-data_analysis}

The experimental $\mu$SR data were analyzed using the following functional form:
\begin{equation}
A(t) = A_{\rm 0,s} P_{\rm s}(t) + A_{\rm 0,bg} P_{\rm bg}(t),
\label{eq:asymmetry}
\end{equation}
where the subscripts s and bg denote the sample and background contributions, respectively, and $A_0$ and $P$ represent the initial asymmetry and the time evolution of the muon-spin polarization.

The sample polarization function $P_{\rm s}(t)$ was further decomposed into magnetic (m) and nonmagnetic (nm) components with respective weights $f_{\rm m}$ and $1-f_{\rm m}$. In experiments performed on the low-background GPS spectrometer \cite{Amato_RSI_2017}, the background contribution was negligible ($A_{\rm 0,bg} \simeq 0$). In high-pressure $\mu$SR experiments conducted on the GPD instrument \cite{Khasanov_HPR_2016, Khasanov_JAP_2022}, the background originates from muons stopping in the pressure-cell walls and accounts for approximately 50\% of the total signal.

In ZF-$\mu$SR experiments, the magnetic and nonmagnetic contributions to $P_{\rm s}(t)$ were modeled as:
\begin{equation}
P_{\rm s,m}(t) = \frac{2}{3} \sum_i f_{{\rm m},i}
e^{-\lambda_{{\rm T},i} t} \cos(\gamma_\mu B_{{\rm int},i} t)
 + \frac{1}{3} e^{-\lambda_{\rm L} t},
\label{eq:incommensurate}
\end{equation}
and
\begin{equation}
P_{\rm s,nm}(t) = \frac{2}{3} (1 - \sigma_{\rm GKT}^2 t^2)
\exp\!\left(-\frac{\sigma_{\rm GKT}^2 t^2}{2}\right)
+ \frac{1}{3}.
\label{eq:GKT}
\end{equation}
Here, $f_{{\rm m},i}$ is the volume fractions of individual magnetic component, satisfying $\sum_i f_{{\rm m},i}=f_{\rm m}$, $\lambda_{{\rm T},i}$ and $\lambda_{\rm L}$ denote the transverse and longitudinal relaxation rates, respectively, $\gamma_\mu = 851.616$~MHz/T is the muon gyromagnetic ratio, $B_{{\rm int},i}$ is the internal field and $\sigma_{\rm GKT}$ is the Gaussian Kubo–Toyabe relaxation rate
The coefficients $2/3$ and $1/3$ arise from powder averaging: in a polycrystalline sample with randomly oriented grains, the local internal magnetic field directions are distributed isotropically with respect to the initial muon-spin polarization. The initial polarization therefore decomposes into transverse and longitudinal components relative to the local field direction, yielding statistical weights of $2/3$ for the transverse (precessing) component and $1/3$ for the longitudinal (nonprecessing) component \cite{Amato-Morenzoni_book_2024, Schenck_book_1985, Yaouanc_book_2011, Blundell_book_2022}.

For WTF-$\mu$SR measurements, the sample response was modeled by considering only the paramagnetic (nonmagnetic) contribution:
\begin{equation}
P_{\rm s}(t)= (1-f_{\rm m})
e^{-\sigma_{\rm WTF}^2 t^2/2}
\cos(\gamma_\mu B_{\rm WTF} t +\phi),
\label{eq:WTF}
\end{equation}
where $\phi$ is the initial phase of the muon-spin ensemble, $B_{\rm WTF}=5$~mT is the applied weak transverse field, and $\sigma_{\rm WTF}$ is the Gaussian relaxation rate of the nonmagnetic component. This expression describes coherent muon-spin precession in $B_{\rm WTF}$ within the paramagnetic fraction of the sample.

\section{Results}\label{sec:Results}

\subsection{Muon-spin rotation experiments at ambient pressure}\label{sec:muSR_ambient-pressure}

Ambient-pressure WTF- and ZF-$\mu$SR experiments were performed using the low-background GPS spectrometer \cite{Amato_RSI_2017}. The measurements were carried out in the so-called Veto mode, which suppresses events from muons that miss the sample region and continue downstream. Owing to the low-background geometry of the GPS instrument, the limited amount of surrounding material, and the relatively large sample size, the background contribution from muons stopping outside the sample is very small, corresponding to $A_{\rm 0,bg} \simeq 0$.

\subsubsection{WTF-$\mu$SR experiments}

Figure~\ref{fig:WTF_ZP}(a) shows the WTF-$\mu$SR time spectra measured in an applied transverse field of $B_{\rm WTF}=5$~mT at $T=10$~K and 200~K. At 200~K, a clear oscillatory signal is observed with an initial asymmetry $A_0 \simeq 0.25$, corresponding to the maximum asymmetry of the GPS spectrometer \cite{Amato_RSI_2017}. In contrast, at $T=10$~K the oscillations are nearly completely suppressed. The disappearance of the oscillatory signal indicates the presence of strong static internal magnetic fields that rapidly depolarize the muon-spin ensemble, signaling the onset of magnetic order.
The solid lines in Fig.~\ref{fig:WTF_ZP}(a) represent fits of Eq.~\ref{eq:asymmetry} to the data, with the sample contribution described by Eq.~\ref{eq:WTF}. The nearly vanishing oscillatory amplitude at low temperature indicates that the sample is fully magnetically ordered.

\begin{figure*}[htb]
\centering
\includegraphics[width=1.0\linewidth]{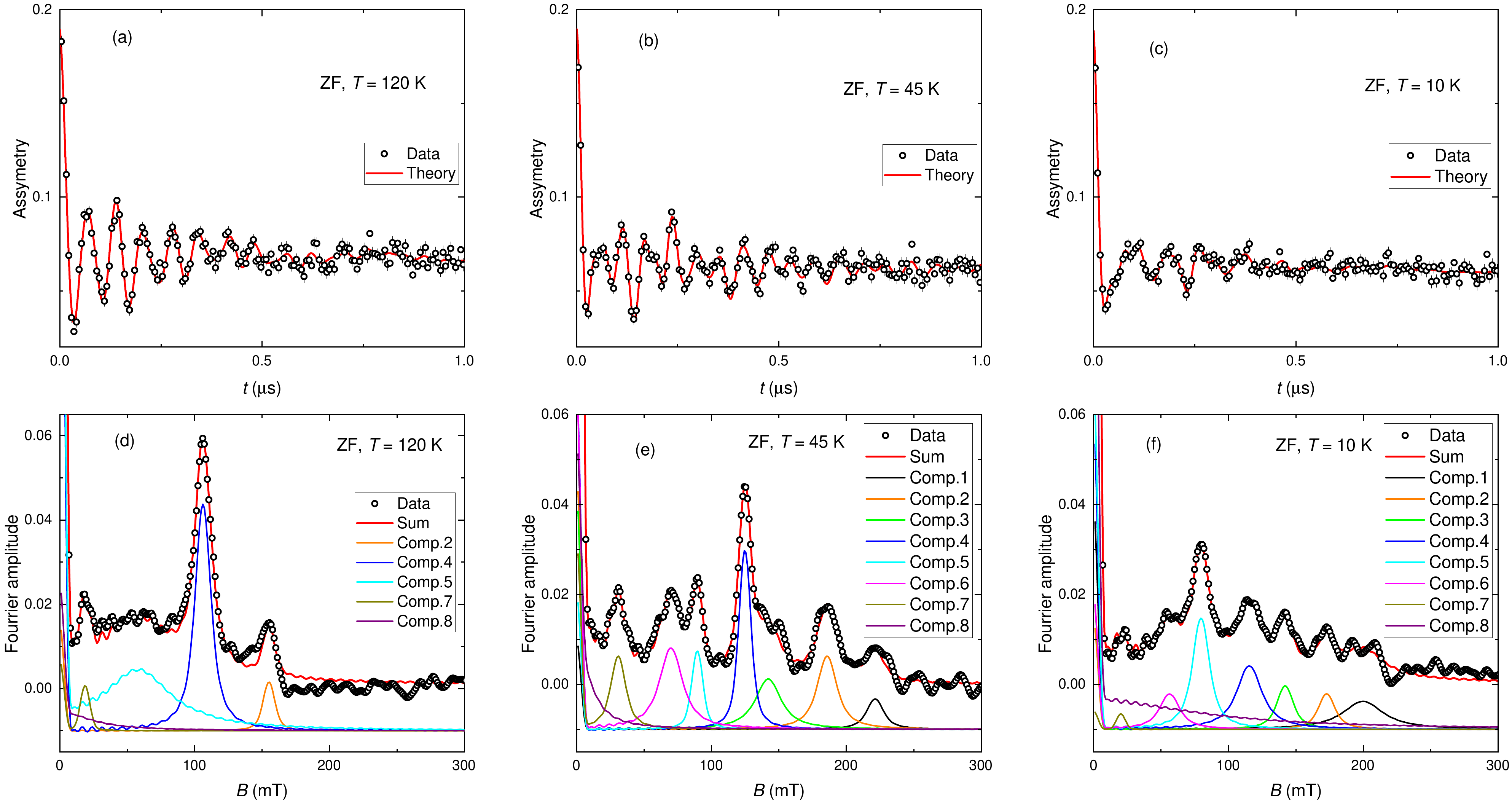}
\caption{%{\bf Zero-field $\mu$SR spectra at $T=120$, 45 and 10~K.}
(a)--(c) ZF-$\mu$SR time spectra measured at $T = 120$~K [panel (a)], 45~K {panle (b)} and $T = 120$~K [panel (c)]. Solid lines represent fits using Eq.~\ref{eq:asymmetry} with the sample contribution described by Eqs.~\ref{eq:incommensurate} and \ref{eq:GKT}. The number of magnetic components: 4 in panels (a) and 5 in panles (b) and (c).
(d)--(f) Fourier transforms of the data in panels (a)--(c). The selection of the components is arbitrary and it is similar for $T = 45$~K and $T = 10$~K data sets, where the components are labeled from 1 to 8 in order of decreasing internal field, i.e. from the highest to the lowest values. For the $T = 120$~K data set, the components are named according to their continuity from the intermediate- to high-temperature region, as follows form Fig.~\ref{fig:Internal-Fields_Volume-Fractions}(a).}
\label{fig:ZF-signals}
\end{figure*}

Figure~\ref{fig:WTF_ZP}(b) displays the temperature dependence of the nonmagnetic volume fraction $1-f_{\rm m}$. Upon cooling, $1-f_{\rm m}$ decreases sharply at the SDW transition, indicating a narrow magnetic transition. The data in the vicinity of the transition were modeled using the phenomenological expression
\begin{equation}
1 - f_{\rm m}(T) =
a\; \left[ 1 + \exp\left(\frac{ T_{\rm m}-T}{\Delta T_{\rm m}}\right) \right]^{-1}
+ c,
\label{eq:Fermi-function}
\end{equation}
where $T_{\rm m}$ and $\Delta T_{\rm m}$ denote the midpoint and width of the magnetic transition, respectively, and $a$ and $c$ represent the amplitude and baseline offset.
The fit yields $T_{\rm SDW}=157.49(4)$~K and $\Delta T_{\rm SDW}=0.65(4)$~K.

Figure~\ref{fig:WTF_ZP}(c) presents the temperature evolution of $1-f_{\rm m}$  measured upon warming and cooling in the vicinity of $T_{\rm SDW}$. A small but clearly resolvable hysteresis is observed, with $T_{\rm SDW}^{\rm warming} > T_{\rm SDW}^{\rm cooling}$. Fits of Eq.~\ref{eq:Fermi-function} to the warming and cooling datasets yield
$T_{\rm SDW}^{\rm warming}=158.23(5)$~K and
$T_{\rm SDW}^{\rm cooling}=157.96(5)$~K, corresponding to a hysteresis width of
$T_{\rm SDW}^{\rm warming}-T_{\rm SDW}^{\rm cooling}=0.27(6)$~K.

Slightly higher values of $T_{\rm SDW}^{\rm warming}$ and $T_{\rm SDW}^{\rm cooling}$, as well as a somewhat larger offset in $1-f_{\rm m}(T)$, were observed for the sample used in the hysteresis measurements. This sample underwent additional annealing (see Sec.~\ref{sec:Sample_preparation}), which may have resulted in a slightly different oxygen content ($\delta$) in Pr$_4$Ni$_3$O$_{10+\delta}$. Given the sensitivity of density-wave transitions to oxygen stoichiometry, such small variations in $\delta$ can account for the minor shift in $T_{\rm SDW}$. The larger offset likely reflects a small residual nonmagnetic fraction and/or a slightly different background contribution.

\subsubsection{ZF-$\mu$SR experiments}\label{sec:ZF-musr-ambient-pressure}

Representative zero-field $\mu$SR time spectra and their corresponding Fourier transforms are shown in Fig.~\ref{fig:ZF-signals} for $T = 120$~K [panels (a) and (d)], $T = 45$~K [panels (b) and (e)], and $T = 10$~K [panels (c) and (f)]. The pronounced oscillatory behavior observed in the time domain demonstrates the presence of static internal magnetic fields at the muon stopping sites, providing clear evidence for long-range magnetic order. The corresponding Fourier transforms are closely related to the internal field distributions and reveal substantial differences among these three characteristic temperatures.

Before discussing the fitting procedure, it should be noted that the magnetic-field distributions shown in Figs.~\ref{fig:ZF-signals}(d)–(f) are highly complex and consist of several distinct peaks. According to neutron-scattering results reported in Ref.~\cite{Samarakoon_PRX_2023}, the magnetic order in Pr$_4$Ni$_3$O$_{10}$ is incommensurate, similar to that established for La$_4$Ni$_3$O$_{10}$ \cite{Zhang_NatCom_2020}. In $\mu$SR studies, such incommensurate magnetic structures are commonly described using Overhauser or shifted-Overhauser field distributions \cite{Amato-Morenzoni_book_2024, Schenck_book_1985, Yaouanc_book_2011, Blundell_book_2022, Overhauser_JPhysChemSolids_1960, Schenck_PRB_2001, Amato_PRB_2014, Khasanov_PRB_MnP_2016}. The Overhauser distribution is characterized by a singularity at $B_{\rm max}$ with a broad plateau extending toward zero field. The shifted-Overhauser distribution exhibits two singularities at $B_{\rm min}$ and $B_{\rm max}$ separated by a broad plateau \cite{Overhauser_JPhysChemSolids_1960, Schenck_PRB_2001, Amato_PRB_2014, Khasanov_PRB_MnP_2016}.

Ideally, a reliable fit model should describe the magnetic structure consistently over the full temperature range. Fits based on sums of Overhauser and shifted-Overhauser components, analogous to the procedure successfully applied to La$_4$Ni$_3$O$_{10}$ in Ref.~\cite{Khasanov_La4310_PRR_2026}, were therefore tested. However, in the present case, neither approach provided fully satisfactory and stable fits across all temperatures. Although improved agreement with the data could be achieved using combinations of shifted-Overhauser components, the resulting solutions were not unique, as different parameter sets yielded comparable fit quality. Consequently, an unambiguous determination of $B_{\rm min}$ and $B_{\rm max}$ for individual components was not possible.

For this reason, a more model-independent approach was adopted, in which the magnetic-field distribution was described as a sum of symmetric Lorentzian lines in the field domain. In the time domain, this corresponds to a sum of cosine oscillations with exponential relaxation (see Eq.~\ref{eq:incommensurate}). While this parametrization does not resolve individual incommensurate components in a microscopic sense, as achieved for La$_4$Ni$_3$O$_{10}$ \cite{Khasanov_La4310_PRR_2026}, it allows the peak positions and their temperature evolution to be tracked in a robust and reproducible manner without overparameterization.

\begin{figure}[htb]
\centering
\includegraphics[width=0.8\linewidth]{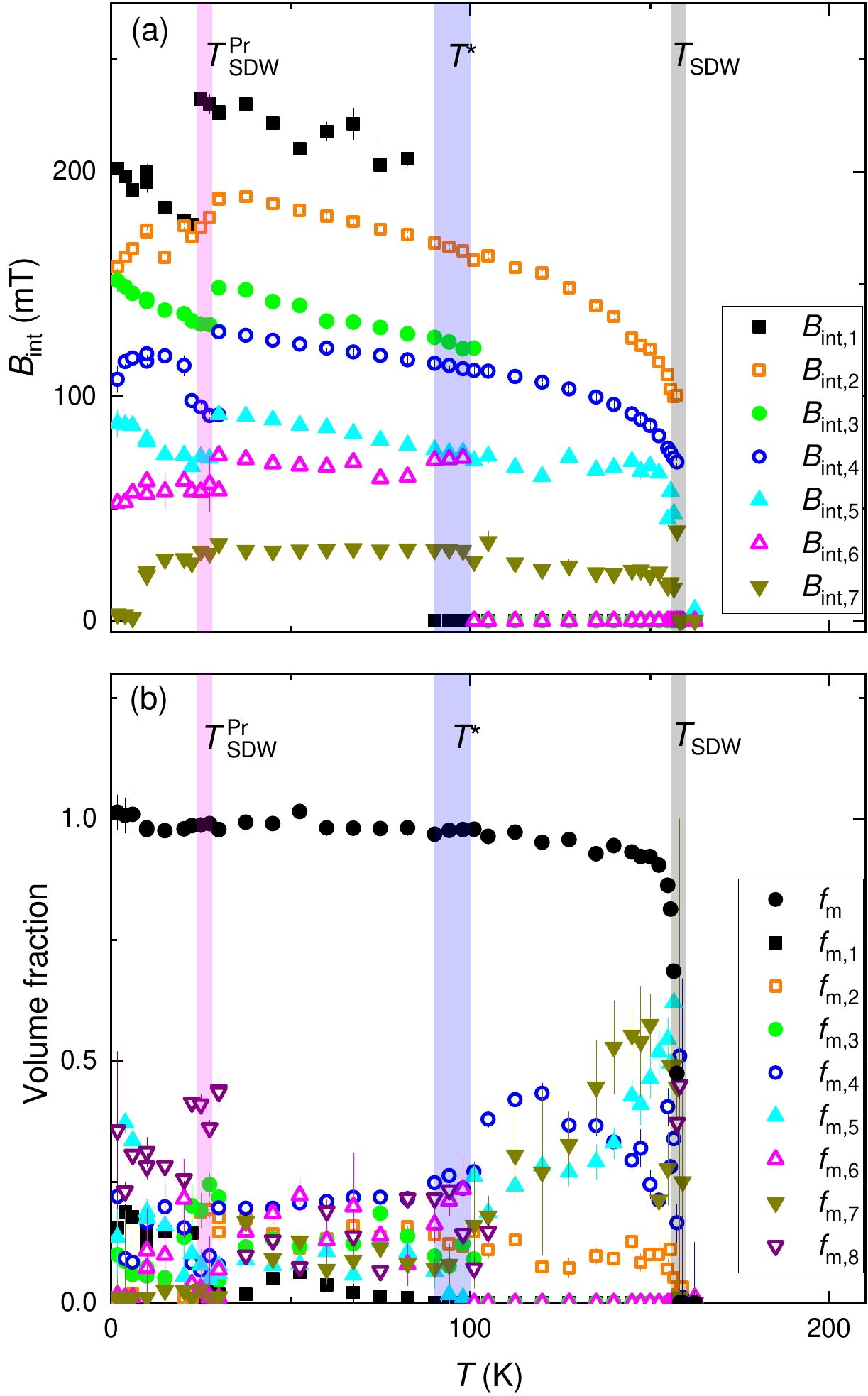}
\caption{
Temperature dependence of (a) the internal fields  $B_{{\rm int},1}$ to $B_{{\rm int},7}$, and (b) the magnetic volume fractions $f_{{\rm m},1}$ to $f_{{\rm m},5}$ together with the total magnetic fraction $f_{\rm m}=\sum_i f_{{\rm m},i}$ obtained from fits to the ZF-$\mu$SR data. Colored stripes indicate the transition temperatures $T_{\rm SDW}$, $T^{\ast}$, and $T_{\rm SDW}^{\rm Pr}$ separating distinct magnetic regimes. The color coding of the individual components is the same as in Fig.~\ref{fig:ZF-signals}.
}
\label{fig:Internal-Fields_Volume-Fractions}
\end{figure}

The solid lines in Fig.~\ref{fig:ZF-signals} represent fits using Eq.~\ref{eq:asymmetry}, with the sample contribution described by Eqs.~\ref{eq:incommensurate} and \ref{eq:GKT}. The selection of the individual internal field components is arbitrary and is identical for the $T = 45$~K and $T = 10$~K data sets [Fig.~\ref{fig:ZF-signals}(e) and (f)], where the components are labeled from 1 to 8 in order of decreasing internal field, i.e., from the highest to the lowest values. At $T = 120$~K [Fig.~\ref{fig:ZF-signals}(d)], the components are named according to their continuity from the intermediate- to high-temperature region, as follows from Fig.~\ref{fig:Internal-Fields_Volume-Fractions}(a). The distribution of magnetic fields can be described as follows:
\\
At $T = 120$~K [Figs.~\ref{fig:ZF-signals}(a) and \ref{fig:ZF-signals}(d)], the spectra are described by four oscillating components (components 2, 4, 5, and 7) and one non-oscillating, fast-relaxing component (component 8). The term ``non-oscillating'' indicates that the corresponding internal field is zero, $B_{{\rm int},8}=0$, so that $\cos(\gamma_{\mu} B_{{\rm int},8} t)=1$, and only exponential relaxation is present.
\\
At $T = 45$~K [Figs.~\ref{fig:ZF-signals}(b) and \ref{fig:ZF-signals}(e)], the field distribution becomes more complex and requires eight components in total. The additional oscillating contributions emerging at this temperature are components 1, 3, and 6. Components 1 and 3 appear as new well-defined peaks, while the initially broad component 5 splits into two distinct contributions. This evolution indicates a modification of the magnetic structure while static magnetic order is preserved.
\\
At $T = 10$~K [Figs.~\ref{fig:ZF-signals}(c) and \ref{fig:ZF-signals}(f)], all seven oscillating and one non-oscillating components remain present. In contrast to the relatively gradual evolution between 120~K and 45~K -- where the principal field positions remain largely unchanged and additional components simply emerge -- the 10~K data exhibit a substantial redistribution of internal fields across nearly all components. This behavior signals a pronounced reconstruction of the magnetic structure at low temperature.

Figure~\ref{fig:Internal-Fields_Volume-Fractions} summarizes the temperature evolution of the internal magnetic fields [$B_{{\rm int,}1}$ to $B_{{\rm int,}7}$, panel (a)] and the magnetic volume fractions [$f_{{\rm m},1}$ to $f_{{\rm m},8}$, and $f_{\rm m}=\sum_i f_{{\rm m},i}$, panel (b)]. Three temperature regions with distinct internal field distributions are clearly visible. The transition temperatures separating these regimes are indicated by colored stripes and occur at $T_{\rm SDW} \simeq 158$~K, $T^{\ast} \sim 90$--100~K, and $T_{\rm SDW}^{\rm Pr} \sim 25$--27~K.

\begin{figure}[htb]
\centering
\includegraphics[width=0.8\linewidth]{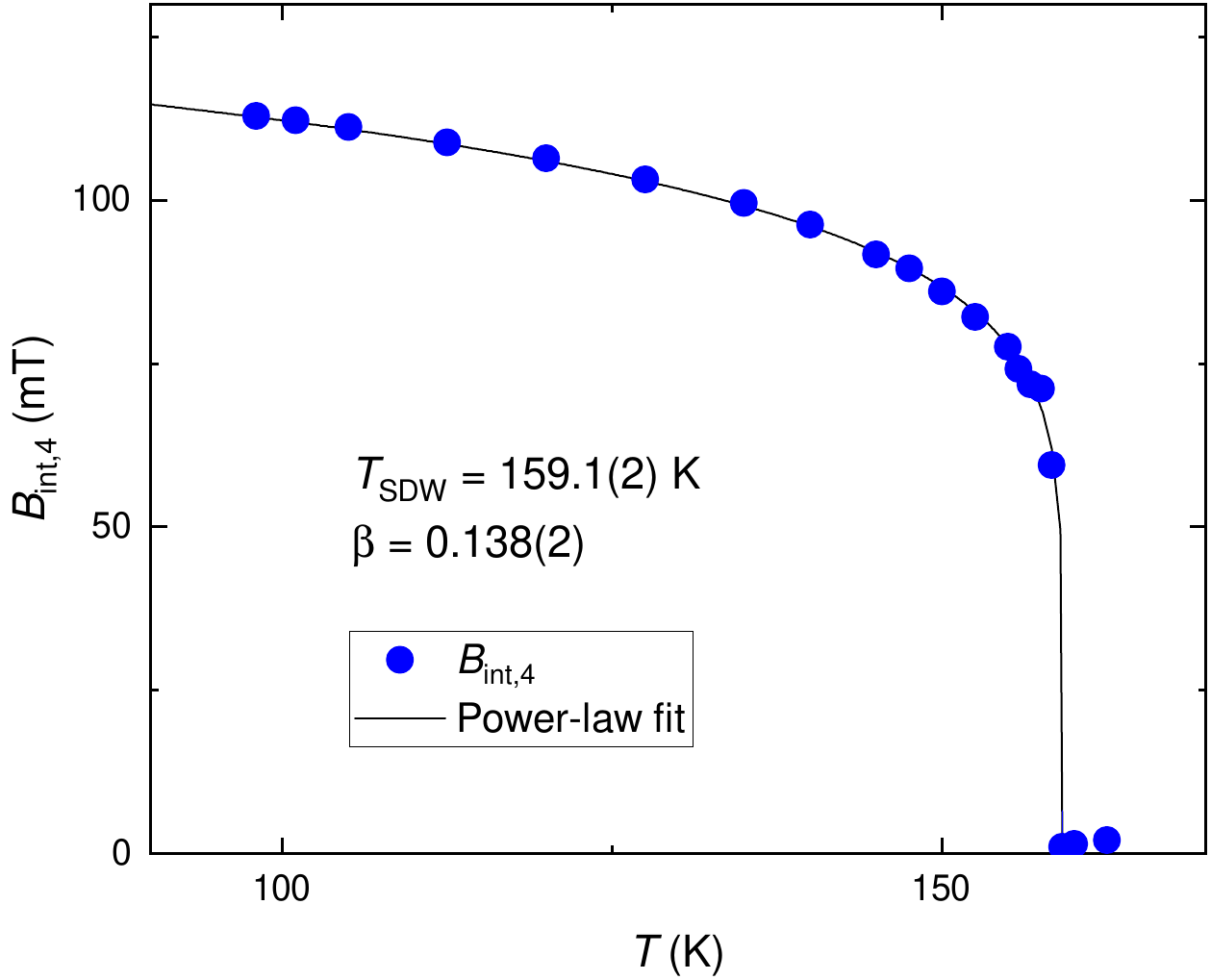}
\caption{
Temperature dependence of the internal field $B_{{\rm int,}4}$ extracted from the ZF-$\mu$SR analysis. The systematic decrease of $B_{{\rm int,}4}$ toward $T_{\rm SDW}$ reflects the suppression of the ordered magnetic moment at the SDW transition.
}
\label{fig:B-internal-2}
\end{figure}

In order to follow the temperature evolution of the internal fields, which are proportional to the ordered magnetic moment, the data in the temperature range between 100 and 160~K were analyzed using a more restricted global fitting procedure. A proportional temperature dependence of the main internal-field components was assumed, namely
$B_{{\rm int},2} \propto B_{{\rm int},4} \propto B_{{\rm int},5} \propto B_{{\rm int},7}$,
with temperature-independent proportionality coefficients.
Component~4, being the most intense contribution, was selected as the reference component.
The temperature evolution of the internal field $B_{{\rm int},4}$ is shown in Fig.~\ref{fig:B-internal-2}.
The systematic decrease of $B_{{\rm int},4}$ upon approaching $T_{\rm SDW}$ reflects the gradual suppression of the ordered magnetic moment toward the SDW transition.

The temperature dependence of $B_{{\rm int},4}$ can be parameterized by the phenomenological power-law expression
\begin{equation}
B_{\rm int}(T) = B_{\rm int}(0)\; \left[1 - \frac{T}{T_{\rm SDW}}\right]^{\beta},
\end{equation}
yielding $T_{\rm SDW}=159.1(2)$ and the effective exponent $\beta = 0.138(2)$.
The comparatively small value of $\beta$ indicates a rapid onset of the ordered moment just below $T_{\rm SDW}$.

\subsection{WTF- and ZF-$\mu$SR experiments under pressure}\label{sec:WTF-ZF-pressure}

In $\mu$SR experiments performed under pressure, a substantial fraction of the muons (approximately 50\% in the present case) stop outside the sample, predominantly in the walls of the pressure cell, and thus contribute to a background signal. As a consequence, the analysis of high-pressure $\mu$SR data requires accurate knowledge of the muon response of the empty pressure cell. This background contribution is typically determined in separate calibration measurements (see, {\it e.g.}, Refs.~\cite{Khasanov_HPR_2016, Khasanov_JAP_2022, Shermadini_HPR_2017, Khasanov_HPR_2022}).

\begin{figure*}[htb]
\centering
\includegraphics[width=1.0\linewidth]{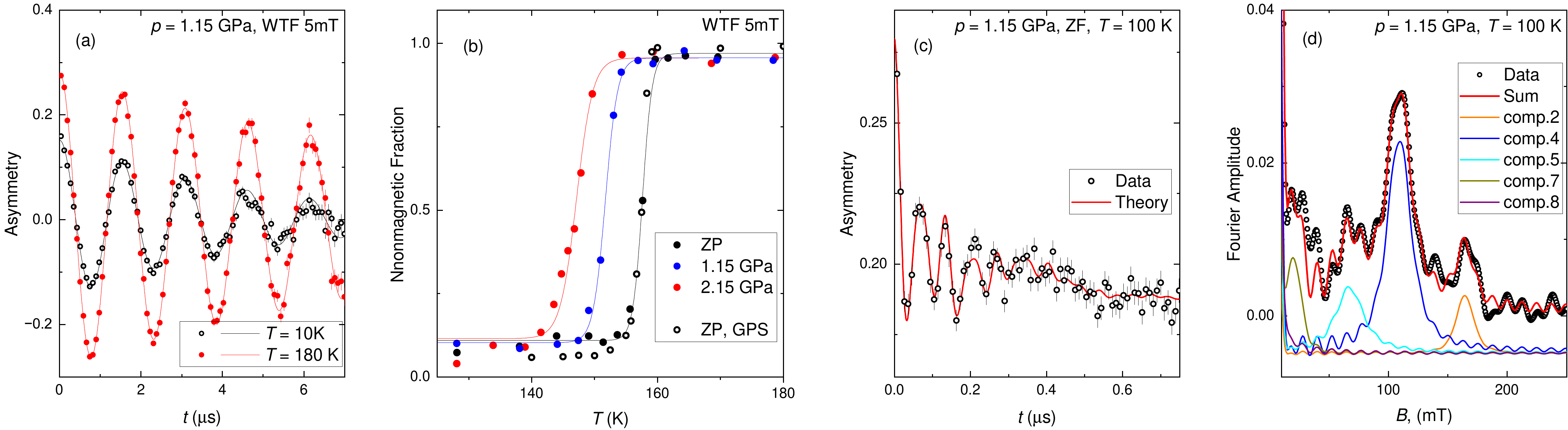}
\caption{
(a) WTF-$\mu$SR time spectra measured at $p \simeq 1.15$~GPa at $T=180$~K and 10~K.
(b) Temperature dependence of the nonmagnetic volume fraction $1-f_{\rm m}$ at $p \simeq 1.15$~GPa. For comparison, the ambient-pressure data obtained with the GPS spectrometer are also shown.
(c) ZF-$\mu$SR time spectrum measured at $p \simeq 1.15$~GPa and $T=100$~K.
(d) Corresponding Fourier transform of the ZF data shown in panel (c). Solid lines represent individual fit components.
}
\label{fig:musr-under-pressure}
\end{figure*}

Figure~\ref{fig:musr-under-pressure}(a) shows the WTF-$\mu$SR time spectra collected at $T=180$~K and $T=10$~K (i.e., above and below the magnetic transition) at a pressure of $p \simeq 1.15$~GPa. In contrast to the ambient-pressure data [see Fig.~\ref{fig:WTF_ZP}(a)], pronounced oscillations remain visible at $T=10$~K. These oscillations originate predominantly from muons stopping in the pressure-cell walls. In the magnetically ordered state, the sample contribution to the oscillatory WTF signal is strongly suppressed, so that the pressure-cell response becomes dominant. The solid lines represent fits using Eq.~\ref{eq:asymmetry}, with the sample contribution described by Eq.~\ref{eq:WTF}.

Figure~\ref{fig:musr-under-pressure}(b) displays the temperature dependence of the nonmagnetic volume fraction $1-f_{\rm m}$ extracted from the fits. For comparison, the corresponding ambient-pressure data measured with the low-background GPS spectrometer, Fig.~\ref{fig:WTF_ZP}(b), are also shown. A clear suppression of the magnetic ordering temperature with increasing pressure is observed. The solid lines represent fits using Eq.~\ref{eq:Fermi-function}.

Zero-field measurements under pressure were performed at $T=10$, 50, and 100~K. The overall behavior is qualitatively similar to that observed in ambient-pressure experiments performed at the GPS instrument (see Fig.~\ref{fig:ZF-signals}). However, due to the more complex line shape, increased background contribution, and reduced counting statistics at the high-pressure GPD spectrometer, it was not possible to resolve the magnetic field distribution with sufficient accuracy at low ($T=10$~K) and intermediate ($T=50$~K) temperatures. Reliable quantitative analysis was achieved only for the high-temperature dataset collected at $T=100$~K.

An example of the ZF-$\mu$SR time spectrum and its corresponding Fourier transform at $p \simeq 1.15$~GPa and $T=100$~K is shown in Figs.~\ref{fig:musr-under-pressure}(c) and \ref{fig:musr-under-pressure}(d). The solid lines in panel~(d) represent the individual fit components, analogous to those identified in Figs.~\ref{fig:ZF-signals}(d)--(f).

\begin{figure}[htb]
\centering
\includegraphics[width=0.8 \linewidth]{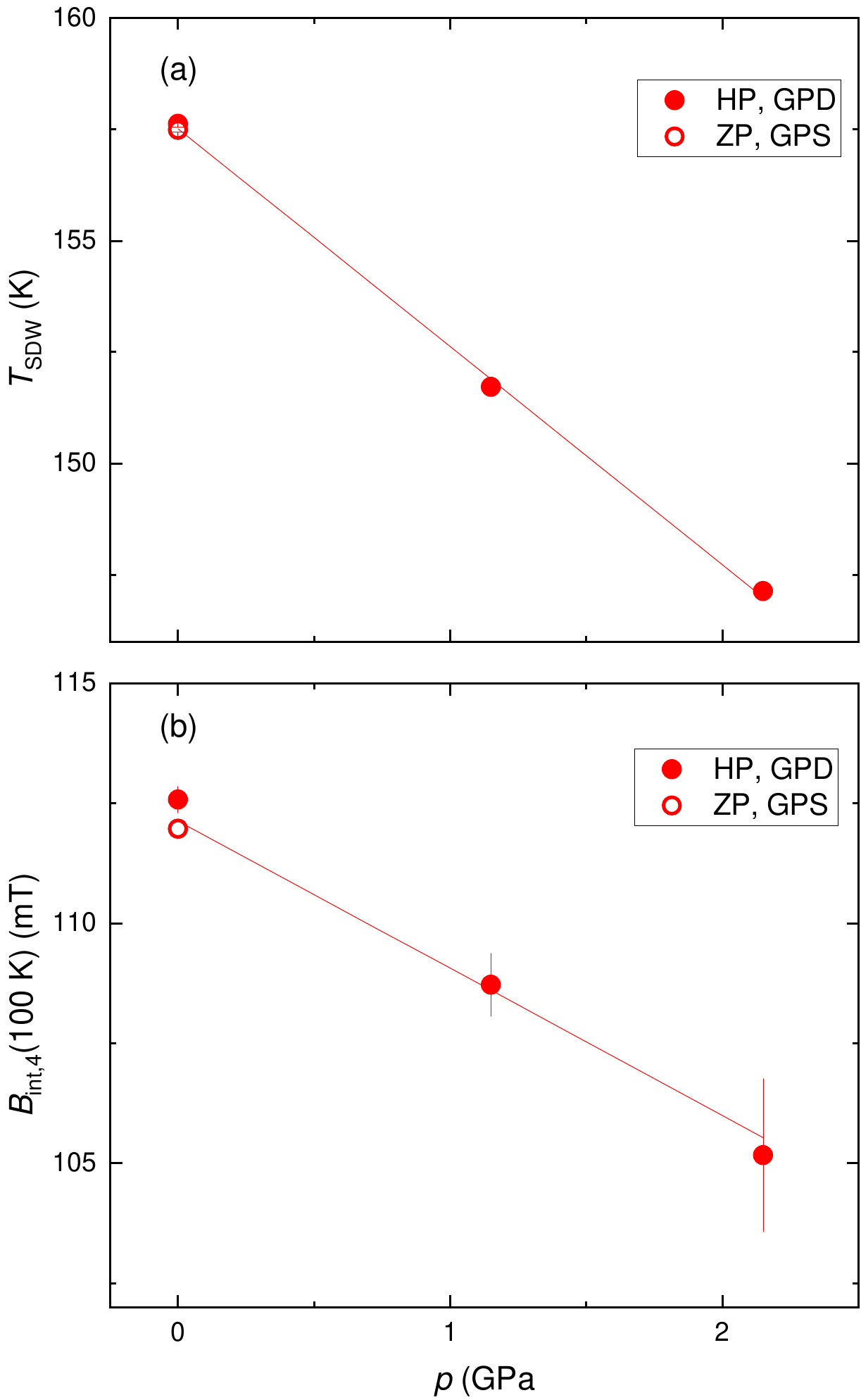}
\caption{
(a) Pressure dependence of the magnetic ordering temperature $T_{\rm SDW}$.
(b) Pressure dependence of the internal field component $B_{{\rm int},4}$ measured at $T = 100$~K.
The labels ``HP, GPD'' and ``ZP, GPS'' correspond to quantities obtained from $\mu$SR measurements under pressure using the GPD spectrometer and at ambient pressure using the GPS muon spectrometer, respectively.
}
\label{fig:pressure-results}
\end{figure}

Figure~\ref{fig:pressure-results} summarizes the pressure evolution of the magnetic properties extracted from the $\mu$SR measurements. Both the magnetic ordering temperature $T_{\rm SDW}$ and the internal field component $B_{{\rm int,}4}(100~\mathrm{K})$ decrease systematically with increasing pressure. Linear fits yield
\begin{align*}
T_{\rm SDW}(p) &= 157.51(8)\,\mathrm{K} - p \cdot 4.9(1)\,\mathrm{K/GPa}, \\
B_{{\rm int,}4}(100{\rm ~K},p) &= 112.1(2)\,\mathrm{mT} - p \cdot 3.1(6)\,\mathrm{mT/GPa}.
\end{align*}

\section{Discussion}\label{sec:Discussions}

\subsection{Multiple magnetic transitions in Pr$_4$Ni$_3$O$_{10}$}

The $\mu$SR measurements reveal three distinct magnetic transitions in Pr$_4$Ni$_3$O$_{10}$. The primary transition at $T_{\rm SDW} \simeq 158$~K marks the onset of static magnetic order associated with an incommensurate spin-density-wave (SDW) state. A second transition at $T^{\ast} \sim 90$--100~K involves a subtle modification of the magnetic structure and is accompanied by the emergence of additional components in the internal magnetic field distribution at the muon stopping sites. A third transition at $T_{\rm SDW}^{\rm Pr} \sim 25$--27~K is characterized by substantial changes in the internal field distribution and is most likely related to the ordering of the Pr $4f$ moments coupled to the Ni sublattice, as stated in Ref.~\cite{Samarakoon_PRX_2023}.

It is important to note that the high-temperature transition at $T_{\rm SDW}$ is clearly identified by both ZF- and WTF-$\mu$SR measurements. In ZF mode, it is manifested by the appearance of spontaneous oscillations in the muon time spectra, indicating the onset of static internal magnetic fields. In WTF mode, it is detected through the reduction of the paramagnetic (i.e., nonmagnetic) asymmetry, reflecting the growth of the magnetically ordered volume fraction. In contrast, the lower-temperature transitions at $T^{\ast}$ and $T_{\rm SDW}^{\rm Pr}$ are resolved only by ZF-$\mu$SR through modifications of the internal field distribution, while the WTF-$\mu$SR data indicate that the sample remains fully magnetically ordered across these transitions.

The observation of magnetic ordering at $T_{\rm SDW}$ together with the additional low-temperature transition at $T_{\rm SDW}^{\rm Pr}$ is consistent with recent reports on Pr$_4$Ni$_3$O$_{10}$ \cite{Samarakoon_PRX_2023}. The present $\mu$SR results confirm and extend these findings by directly resolving the internal magnetic fields and quantitatively determining the internal magnetic field components associated with the respective transitions.

In addition to these two previously reported transitions, our data reveal a clear anomaly at $T^{\ast} \sim 90$--100~K. This intermediate transition is characterized by a redistribution of the internal magnetic fields without a significant change in the total magnetic volume fraction. Such behavior indicates a reconstruction or reorientation within the existing SDW phase rather than the formation of a distinct new magnetic state. A similar intermediate-temperature transition has recently been detected by $\mu$SR in La$_4$Ni$_3$O$_{10}$ \cite{Khasanov_La4310_PRR_2026, Cao_PRB_2025}, where it was attributed to a modification of the magnetic structure within the SDW phase. Related behavior has also been reported in the alternating monolayer–trilayer compound La$_3$Ni$_2$O$_7$ \cite{Khasanov_La327-1313_Arxiv_2025}, suggesting a common mechanism in layered nickelates containing trilayer structural units.

The overall consistency between our observations and these recent studies suggests that the presence of multiple magnetic transitions -- namely, the onset of SDW order, an intermediate reconstruction within the SDW phase, and a low-temperature rare-earth driven transition -- may represent a generic feature of trilayer and structurally related Ruddlesden-Popper nickelates.

\subsection{Character of the high-temperature SDW transition}\label{sec:Nature-of-the-SDW}

The combined WTF- and ZF-$\mu$SR measurements provide detailed insight into the character of the SDW transition. In the WTF configuration, the nonmagnetic volume fraction $1 - f_m(T)$ was determined as a function of temperature for both increasing and decreasing temperature runs as presented in Fig.~\ref{fig:WTF_ZP}~(c). A finite thermal hysteresis of $\Delta T_{\rm hyst} = 0.27(6)$~K is obtained, with $T_{\rm SDW}^{\rm warming} > T_{\rm SDW}^{\rm cooling}$. The fact that the transition temperature determined upon warming exceeds that obtained upon cooling reflects metastability near the transition. In a first-order–like scenario, the ordered phase can persist upon warming (superheating), while the paramagnetic phase can persist upon cooling (supercooling), resulting in a finite hysteresis interval \cite{Landau_Lifshitz_StatPhys, Binder_RPP_1987, Imry_Wortis_PRB_1979,Khasanov_SciRep_2015}.

Complementary information is provided by ZF-$\mu$SR measurements. Below $T_{\rm SDW}$, spontaneous muon-spin precession develops, demonstrating the formation of static internal magnetic fields and long-range magnetic order. The internal field increases rapidly upon cooling, reflecting the sharp establishment of the ordered state (see Fig.~\ref{fig:B-internal-2}).
In general, the temperature evolution of an order parameter is described by a phenomenological expression of the form: $ B_{\rm int}(T) = B_{\rm int}(0)\; \left[1 - \left(T/T_{\rm SDW}\right)^{\alpha}\right]^{\beta}$,
where the exponent $\alpha$ governs the overall temperature evolution over a broad range, while $\beta$ characterizes the critical behavior in the immediate vicinity of the transition. In the present analysis, a simplified form without the additional exponent $\alpha$ was employed (see Eq.~\ref{eq:Fermi-function}), since the primary interest lies in the behavior close to $T_{\rm SDW}$, where $\beta$ determines the curvature of the order parameter and $\alpha$ becomes less relevant.

It is further important to note that the fit of $B_{\rm int}(T)$ was restricted to the temperature interval between 100~K and $T_{\rm SDW}$, where the magnetic-field distribution exhibits its simplest structure (see Fig.~\ref{fig:ZF-signals}). At lower temperatures, the internal-field distribution becomes more complex due to additional magnetic components, making a single-order-parameter description less well justified.

A power-law fit, as presented in Sec.~\ref{sec:ZF-musr-ambient-pressure}, yields the effective critical exponent $\beta = 0.138(2)$. This value is close to the theoretical prediction $\beta_{\rm 2D}=1/8$ for the two-dimensional (2D) Ising universality class \cite{Onsager1944, BaxterBook}, and is substantially smaller than typical values for three-dimensional (3D) magnetic transitions (e.g., $\beta \simeq 0.33$ for 3D Ising and $\beta \simeq 0.36$ for 3D Heisenberg systems) \cite{PelissettoVicari2002}.

From a structural perspective, Pr$_4$Ni$_3$O$_{10}$ consists of trilayer NiO$_2$ blocks separated by rock-salt-type spacer layers \cite{Pei_arxiv_2024, Huangfu_PRB_2020, Zhang_PRM_2020, Zhang_arxiv_2025, Samarakoon_PRX_2023, Tsai_JSSC_2020, Pei_JACS_2026}. Strong magnetic coupling is expected within each trilayer unit, whereas inter-block coupling is reduced due to spatial separation \cite{Zhang_NatCom_2020, Samarakoon_PRX_2023, Khasanov_La327-1313_Arxiv_2025, Khasanov_La4310_PRR_2026}. This hierarchy of exchange interactions promotes quasi-two-dimensional magnetic correlations over an extended temperature range. The pronounced sharpness of the SDW onset and the small effective exponent $\beta$ are consistent with this quasi-two-dimensional magnetic character.

However, in view of the finite hysteresis extracted from the fits to $1 - f_m(T)$, the exponent $\beta$ should be regarded as an effective phenomenological parameter describing the steepness of the temperature evolution rather than a universal critical exponent of a strictly continuous phase transition. The coexistence of a sharp order-parameter onset and a small but finite hysteresis suggests that the SDW transition lies close to a first-order instability. In trilayer Ruddlesden--Popper nickelates like La$_4$Ni$_3$O$_{10}$ and Pr$_4$Ni$_3$O$_{10}$, SDW order is intertwined with CDW order \cite{Zhang_NatCom_2020, Samarakoon_PRX_2023, Jia_PRX_2026}, and coupling between spin and charge order parameters may enhance discontinuous tendencies. Such coupled spin–charge dynamics provide a plausible microscopic origin for the weakly first-order--like character observed here \cite{Norman_PRB_2025}.

\subsection{Pressure evolution of the SDW state}

Application of hydrostatic pressure leads to a systematic suppression of the SDW transition temperature (see Fig.~\ref{fig:pressure-results}). The pressure coefficient ${\rm d}T_{\rm SDW}/{\rm d}p = -4.9(1)$~K/GPa is obtained from the analysis.

It is instructive to compare this behavior with that reported for La$_4$Ni$_3$O$_{10}$, where a significantly stronger suppression of the magnetic ordering temperature under pressure has been observed, ${\rm d}T_{\rm SDW}/{\rm d}p \simeq -13(1)$~K/GPa \cite{Khasanov_La4310_PRR_2026}. The more rapid reduction of $T_{\rm SDW}$ in the La compound is also reflected in the pressure scale at which superconductivity emerges. In La$_4$Ni$_3$O$_{10}$, superconductivity appears above $\sim 15$~GPa, whereas in Pr$_4$Ni$_3$O$_{10}$ it sets in only at higher pressures of approximately 20--25~GPa \cite{Zhu_Nature_2024, Zhang_arxiv_2023, Xu_NatCom_2025, Pei_arxiv_2024, Huang_CPL_2024, Pei_JACS_2026}.
The comparatively weaker suppression of magnetism in Pr$_4$Ni$_3$O$_{10}$ is therefore consistent with its larger critical pressure for the onset of superconductivity. This suggests that the robustness of magnetic order against lattice compression plays an important role in determining the pressure scale at which superconductivity develops in trilayer nickelates.

In addition to lowering $T_{\rm SDW}$, pressure reduces the internal magnetic field, indicating a decrease of the ordered Ni moment. Since the internal field sensed by the muons is proportional to the ordered moment, this behavior demonstrates that pressure weakens not only the stability of the SDW phase but also the magnitude of the magnetic order parameter itself. Such trends are consistent with enhanced electronic bandwidth and reduced nesting conditions under compression, which diminish the SDW instability. The present results thus provide microscopic insight into the interplay between magnetism and superconductivity in trilayer Ruddlesden--Popper nickelates.

A quantitative estimate of the intrinsic pressure dependence of the ordered Ni moment can be obtained from the pressure evolution of the internal field. Since the internal magnetic field at the muon stopping site is proportional to the ordered moment $M$, one may write ${\rm d}B_{\rm int}/{\rm d}p \propto {\rm d}M/{\rm d}p$. However, the ZF-$\mu$SR measurements under pressure were performed at a fixed temperature ($T = 100$~K), while $T_{\rm SDW}$ decreases with increasing pressure. Consequently, the reduced temperature $T/T_{\rm SDW}(p)$ increases with pressure, leading to an additional suppression of the measured internal field unrelated to the intrinsic reduction of the magnetic order-parameter amplitude.

To correct for this effect, we employ the phenomenological power-law temperature dependence of the order parameter described in Sec.~\ref{sec:ZF-musr-ambient-pressure}, assuming that the exponent $\beta$ remains pressure independent. A straightforward evaluation shows that the reduced-temperature contribution at 100~K amounts to $ -8(1)\times10^{-3}$~GPa$^{-1}$. From the measured slope ${\rm d}\ln B_{{\rm int,}4}/{\rm d}p = -2.8(5)\times10^{-2}$~GPa$^{-1}$, the intrinsic pressure dependence of the order-parameter amplitude is therefore estimated as ${\rm d}\ln M / {\rm d}p={\rm d}\ln B_{\rm int} / {\rm d}p = -2.0(5)\times10^{-2}$~GPa$^{-1}$. Within experimental uncertainty, this value is very close to that reported for La$_4$Ni$_3$O$_{10}$, where ${\rm d}\ln M / {\rm d}p = -1.8(5)\times10^{-2}$~GPa$^{-1}$
was found \cite{Khasanov_La4310_PRR_2026}. This similarity suggests that the intrinsic pressure weakening of the ordered Ni moment is a common feature of trilayer Ruddlesden–Popper nickelates.

Taken together, these results indicate that pressure weakens the SDW state not only by lowering $T_{\rm SDW}$ but also by moderately reducing the magnitude of the ordered Ni moment itself. This demonstrates that compression affects both the thermodynamic stability of the SDW phase and the amplitude of the magnetic order parameter. The suppression of magnetism under pressure is thus intrinsic and cannot be attributed solely to the shift of the transition temperature, but instead reflects a genuine weakening of the static magnetic state.

\section{Conclusions}\label{sec:Conclusions}

We have performed a comprehensive $\mu$SR investigation of the trilayer Ruddlesden-Popper nickelate Pr$_4$Ni$_3$O$_{10}$ at ambient pressure and under hydrostatic pressure up to 2.2~GPa. At ambient pressure, three distinct magnetic transitions are identified: the onset of spin-density-wave (SDW) order at $T_{\rm SDW} \simeq 158$~K, an intermediate transition at $T^{\ast} \simeq 90$--100~K associated with a subtle modification of the magnetic structure, and a low-temperature transition at $T_{\rm SDW}^{\rm Pr} \simeq 25$--27~K attributed to ordering of the Pr sublattice.

The primary SDW transition is characterized by a sharp development of static magnetism and a narrow transition width $\Delta T_{\rm SDW} \simeq 0.65(4)$~K. Weak-transverse-field measurements reveal a finite thermal hysteresis of $\simeq 0.27(6)$~K, with $T_{\rm SDW}^{\rm warming} > T_{\rm SDW}^{\rm cooling}$, indicating metastability and weakly first-order--like behavior. Although the temperature dependence of the $\mu$SR internal field can be parameterized by a power-law form, the presence of hysteresis suggests that the transition lies close to a first-order boundary rather than representing a strictly continuous instability.

The trilayer crystal structure, consisting of three strongly coupled NiO$_2$ planes forming a single block and separated from neighboring blocks by rock-salt layers, provides a natural basis for anisotropic magnetic behavior. While the magnetic coupling is strong within each trilayer unit, it is substantially reduced between adjacent trilayer blocks. The sharp SDW onset and the observed hysteresis point to a delicate balance between competing ordered states, likely involving intertwined spin and charge degrees of freedom.

Under applied pressure, both the SDW transition temperature and the ordered Ni magnetic moment decrease linearly. The suppression rate of the transition temperature amounts to ${\rm d}T_{\rm SDW}/{\rm d}p = -4.9(1)$~K/GPa, while the ordered moment is reduced at a rate of ${\rm d}\ln M/{\rm d}p = -2.0(5)\times10^{-2}$~GPa$^{-1}$ (corresponding to about $2\%$ per GPa), as inferred from the pressure dependence of the $\mu$SR internal field. This behavior demonstrates a gradual weakening of the SDW instability under compression.
%The comparatively moderate pressure sensitivity of the ordered moment is consistent with the higher critical pressure required for superconductivity in Pr$_4$Ni$_3$O$_{10}$ relative to its La analogue.

Overall, our results establish the magnetic phase diagram of Pr$_4$Ni$_3$O$_{10}$ in the pressure range below the superconducting regime and provide a quantitative microscopic baseline for understanding the evolution of density-wave order and magnetism in trilayer RP nickelates.

\section*{ACKNOWLEDGEMENTS}
Z.G. acknowledges support from the Swiss National Science Foundation (SNSF) through SNSF Starting Grant (No. TMSGI2${\_}$211750). I.P. acknowledges financial support from Paul Scherrer Institute research grant No. 2021\_0134. D.J.G. acknowledges support from the Swiss National Science Foundation (SNSF) through Grant No. 200021E\_238113.

%\ack{Z.G. acknowledges support from the Swiss National Science Foundation (SNSF) through SNSF Starting Grant (No. TMSGI2${\_}$211750). I.P. acknowledges financial support from Paul Scherrer Institute research grant No. 2021\_0134. D.J.G. acknowledges support from the Swiss National Science Foundation (SNSF) through Grant No. 200021E\_238113.}

%\roles{R.K. conceived the project, performed $\mu$SR experiments, analysed the data and wrote the manuscript. D.J.G. and E.P. synthesized the sample and performed x-ray and thermogramme experiments. I.P. and V.P. performed neutron powder diffraction measurement. Z.G., T.J.H., D.J.G., and H.L. discussed the manuscript and gave critical suggestions.}

%\data{The data that support the findings of this study are available upon reasonable request from the corresponding author.}

\end{document}